\documentclass[fleqn,usenatbib]{mnras}

\usepackage{mathrsfs}

\usepackage[T1]{fontenc}
\usepackage{ae,aecompl}
\usepackage{booktabs}

\usepackage{graphicx}	
\usepackage{amsmath}	
\usepackage{amssymb}	
\usepackage{float}

\usepackage{empheq}
\usepackage{newtxtext,newtxmath}

\title[Revealing the reionisation with future FRBs]{Revealing the cosmic reionisation history with fast radio bursts in the era of Square Kilometre Array}

\author[T. Hashimoto et al.]{
Tetsuya Hashimoto,$^{1,2}$\thanks{E-mail: tetsuya@phys.nthu.edu.tw}
Tomotsugu Goto,$^{1}$
Ting-Yi Lu,$^{1}$
Alvina Y. L. On,$^{1,2,3}$
\newauthor
Daryl Joe D. Santos,$^{1}$
Seong Jin Kim, $^{1}$
Ece Kilerci-Eser,$^{4}$
Simon C.-C. Ho,$^{1}$
\newauthor
Tiger Y.-Y. Hsiao,$^{1}$
and Leo Y.-W. Lin$^{5}$
\\
$^{1}$Institute of Astronomy, National Tsing Hua University, 101, Section 2. Kuang-Fu Road, Hsinchu, 30013, Taiwan (R.O.C.)\\
$^{2}$Centre for Informatics and Computation in Astronomy (CICA), National Tsing Hua University, 101, Section 2. Kuang-Fu Road, Hsinchu, 30013, Taiwan (R.O.C.)\\
$^{3}$Mullard Space Science Laboratory, University College London, Holmbury St Mary, Surrey RH5 6NT, UK\\
$^{4}$Sabanc{\i} University, Faculty of Engineering and Natural Sciences, 34956, Istanbul, Turkey\\
$^{5}$Department of Physics, National Tsing Hua University, 101, Section 2. Kuang-Fu Road, Hsinchu, 30013, Taiwan (R.O.C.)
}

\date{Accepted 2021 January 18. Received 2020 December 24; in original form 2020 October 22}

\pubyear{2021}

\begin{document}
\label{firstpage}
\pagerange{\pageref{firstpage}--\pageref{lastpage}}
\maketitle

\begin{abstract} 
Revealing the cosmic reionisation history is at the frontier of extragalactic astronomy.
The power spectrum of the cosmic microwave background (CMB) polarisation can be used to constrain the reionisation history. 
Here we propose a CMB-independent method using fast radio bursts (FRBs) to directly measure the ionisation fraction of the intergalactic medium (IGM) as a function of redshift.
FRBs are new astronomical transients with millisecond timescales. 
Their dispersion measure (DM$_{\rm IGM}$) is an indicator of the amount of ionised material in the IGM.
Since the differential of DM$_{\rm IGM}$ against redshift is proportional to the ionisation fraction, our method allows us to directly measure the reionisation history without any assumption on its functional shape.
As a proof of concept, we constructed mock non-repeating FRB sources to be detected with the Square Kilometre Array, assuming three different reionisation histories with the same optical depth of Thomson scattering.
We considered three cases of redshift measurements: (A) spectroscopic redshift for all mock data, (B) spectroscopic redshift for 10\% of mock data, and (C) redshift estimated from an empirical relation of FRBs between their time-integrated luminosity and rest-frame intrinsic duration.
In all cases, the reionisation histories are consistently reconstructed from the mock FRB data using our method.
Our results demonstrate the capability of future FRBs in constraining the reionisation history.
\end{abstract}

\begin{keywords}
(cosmology:) dark ages, reionization, first stars -- (cosmology:) early Universe -- radio continuum: transients -- stars: magnetars -- stars: neutron -- stars: luminosity function, mass function
\end{keywords}



\section{Introduction}
\label{introduction}
The cosmic (hydrogen) reionisation is the latest phase transition of the Universe from neutral to ionised state \citep[e.g.,][]{Barkana2001}.
During the reionisation epoch, the intergalactic medium was presumably reionised by strong radiation from the first stars and the first galaxies \citep[e.g.,][]{Bouwens2012,Finkelstein2012,Duncan2015,Robertson2015,Naidu2020} with probably minor contributions from quasars (QSOs) \citep[e.g.,][]{Onoue2017,Yoshiura2017,Hassan2018,Matsuoka2018}.
Revealing how and when the reionisation happened is at the forefront of extragalactic astronomy.

Several methods have been proposed to investigate the cosmic reionisation.
Using Ly$\alpha$ absorption and its damping wing is one of the traditional methods.
High-redshift (high-$z$) bright sources such as QSOs and optical afterglows of long gamma-ray bursts (LGRBs) are used as background light sources \citep[e.g.,][]{Fan2002,Goto2006,Becker2015,Barnett2017,Bosman2018,TYLu2020,Totani2014,Hartoog2015,Melandri2015}.
Their light is absorbed by the intervening intergalactic medium between the source and the observer, producing Ly$\alpha$ absorption features which are imprinted in their spectra.
This allows us to measure the ionisation fraction in the intergalactic medium as a function of redshift.
This method does not assume any model of the reionisation in calculating the ionisation fraction from the observed Ly$\alpha$ absorption.
The tail end of cosmic reionisation is largely confirmed to be $z\sim6$ by high-redshift QSOs and LGRB afterglows in previous studies \citep[e.g.,][]{Fan2002,Fan2006,Totani2006}.
However, applying this method beyond $z\sim7$ is challenging because the number densities of bright high-$z$ QSOs and LGRBs rapidly decline towards higher redshifts \citep[e.g.,][]{Matsuoka2018,Wang2019,Perley2016}.
In addition, the observed spectra require both high spectral resolution and high signal-to-noise ratio, given that the transmitted light is very faint due to the strong Ly$\alpha$ absorption.
This observational difficulty hampers accurate measurements of ionisation fraction at $z\gtrsim7$, placing relatively large uncertainties \citep[e.g.,][]{Banados2018,Yang2020}.

On the other hand, cosmic microwave background (CMB) observations can reach high redshifts up to the recombination epoch of the Universe (i.e., $z\sim1100$) beyond QSOs and LGRBs \citep[e.g.,][]{Hinshaw2013,Planck2020}.
CMB allows us to measure the optical depth of Thomson scattering of free electrons in the intergalactic medium ($\tau_{\rm CMB}$).
Since $\tau_{\rm CMB}$ is an integration over redshift from $z=0$ to $\sim1100$, it also depends on the epoch of reionisation.
Recently, the optical depth was estimated to be $\tau_{\rm CMB}=0.059\pm0.006$ from {\it Planck} observations \citep{Pagano2020}.
This value corresponds to an averaged reionisation redshift at $z=8.14\pm0.61$, assuming a Tanh reionisation history \citep{Pagano2020}.
This estimate is largely consistent with the end of reionisation epoch at $z\sim6$ as revealed by QSOs and LGRBs \citep[e.g.,][]{Fan2002,Fan2006,Totani2006}.
However, solely using $\tau_{\rm CMB}$ is not sensitive to the detailed ionisation history, i.e., the ionisation fraction as a function of redshift, because it measures only the amount of electron scattering integrated over redshift.
The shape of the power spectrum of CMB polarisation encodes information on the detailed redshift evolution of the ionisation fraction \citep[e.g.,][]{Zaldarriaga1997,Kaplinghat2003,Planck2020}.
Therefore, the CMB polarisation in principle can constrain the early reionisation at $z\sim15$-30.
The optical depth between $z=15$ and $z=30$ is constrained to $\tau_{\rm CMB} (15,30) < 0.007$ \citep{Planck2020}.

21-cm absorption or emission features of atomic hydrogen in an all-sky radio spectrum is also a powerful tracer of cosmic reionisation \citep[e.g.,][]{Furlanetto2006,Pritchard2012,Barkana2016}.
The absorption feature appears when the 21-cm spin temperature is lower than the CMB temperature in the early Universe.
This condition is satisfied when gas temperature is coupled with the 21-cm spin temperature due to resonant scattering of Ly$\alpha$ photons emitted from the first stars and galaxies \citep[e.g., ][]{Bowman2018}.
Over time, the gas and coupled 21-cm spin temperatures are expected to become higher than the CMB temperature due to the heating effects by galaxies and active galactic nuclei, ending the absorption feature.
Therefore, the 21-cm absorption is sensitive to the beginning of cosmic reionisation before the majority of the intergalactic medium starts to become reionised.
\citet{Bowman2018} claimed the detection of the 21-cm absorption around $z\sim17$ from observations with the Experiment to Detect the Global Epoch of Reionization Signature (EDGES) low-band instrument \citep{Bowman2008,Rogers2008}.
The absorption feature spans from $z\sim15$ to $\sim20$.
The $z\sim20$ and $z\sim15$ edges of the absorption feature correspond to the redshift at which the 21-cm spin temperature became lower than the CMB temperature and the redshift at which the gas temperature exceeded the CMB temperature, respectively.
This implies that cosmic reionisation has most likely started at $z\lesssim20$.
However, we caution that the EDGES observation has large measurement errors. Therefore, the actual redshifts at which the signal crosses zero might be very different from the abrupt drops. In addition, the EDGES signal might be non-cosmological, in which case there is no associated redshift.
During reionisation, the 21-cm spin temperature becomes much higher than the CMB temperature, which causes the 21-cm emission feature.
The 21-cm emission directly traces the ionisation fraction in the intergalactic medium \citep[e.g.,][]{Bowman2010}.
However, no clear detection of the 21-cm emission in the intergalactic medium at the reionisation epoch has been reported yet.

None of these methods has provided a complete picture of the cosmic reionisation history yet.
In contrast, fast radio bursts (FRBs) have a potential to allow us to overcome the difficulties mentioned above.
FRB is a new type of astronomical transients, which is a bright burst in radio (order of Jy) with a millisecond time scale \citep[e.g.,][]{Lorimer2007}.
A unique observable of FRBs is the dispersion measure (DM), which is characterised by a time lag of burst arrival depending on observed frequencies.
The DM is an indicator of the amount of ionised material between the source of FRB and the observer.
The majority of FRBs shows an observed DM (DM$_{\rm obs})\gtrsim 300$ pc cm$^{-3}$, indicating that the sources must be extragalactic \citep[e.g.,][]{Cordes2019,Hashimoto2020}.
Since the DM is sensitive to the ionised intergalactic medium, some studies \citep[e.g., ][]{Caleb2019,KitLau2020,Linder2020} proposed to use FRBs to constrain the \lq helium reionisation\rq\ at $z\sim3$-4 \citep[e.g.,][]{Becker2011} with future FRBs.
However, the sample size and redshift range of current FRB data are too small to constrain the reionisation history \citep[e.g.,][]{Petroff2016}.
The Square Kilometre Array (SKA) is expected to detect hundreds to thousands of FRBs at $z\gtrsim6$ \citep[e.g.,][]{Fialkov2017,Hashimoto2020b}, which motivates studies on the hydrogen reionisation at $z\gtrsim6$ using future DM measurements \citep[e.g.,][]{Ioka2003,Inoue2004,Fialkov2016,Dai2020}.

FRBs provide a new method to measure the amount of ionised intergalactic medium independent of the CMB.
The CMB has only one source redshift and thus can only probe the ionisation fraction along the line of sight all the way out to $z\sim1100$. 
However, if high-$z$ FRBs exist, they are sources at multiple redshifts and can probe the ionised fraction out to different redshifts \citep[e.g.,][]{Ioka2003, Inoue2004,Fialkov2016}.

\citet{Zheng2014} pointed out that the differential of averaged DM$_{\rm IGM}$ against redshift (i.e., $d\langle{\rm DM_{\rm IGM}}\rangle/dz$) is more sensitive to the reionisation history than the DM$_{\rm IGM}$ itself.
Motivated by the previous work, we propose to use $d\langle{\rm DM_{\rm IGM}}\rangle/dz$, which will be available for future FRBs. 
The $d\langle{\rm DM_{\rm IGM}}\rangle/dz$ allows us to directly measure the ionisation fraction without any assumption on a functional shape of the reionisation history.

The structure of this paper is as follows. 
In Section \ref{method}, we describe the ionisation fraction as a function of redshift derived from $d\langle{\rm DM_{\rm IGM}}\rangle/dz$.
We generate mock FRB data to be detected with the SKA in Section \ref{mock}.
In Section \ref{results}, we show ionisation fractions as a function of redshift reconstructed from the mock FRB data, followed by discussions in Section \ref{discussions} and conclusions in Section \ref{conclusions}.
Throughout this paper, we assume the {\it Planck15} cosmology \citep{Planck2016} as a fiducial model, i.e., $\Lambda$ cold dark matter cosmology with ($\Omega_{m}$,$\Omega_{\Lambda}$,$\Omega_{b}$,$h$)=(0.307, 0.693, 0.0486, 0.677), unless otherwise mentioned.

\section{Differential of dispersion measure}
\label{method}
The $\tau_{\rm CMB}$ and DM$_{\rm IGM}$ can be measured from the CMB and FRBs, respectively.
Both quantities integrate the electron density ($n_{\rm e}$) in the intergalactic medium as follows.
\begin{equation}
\label{eqtauCMB}
\tau_{\rm CMB} = \int_{z=0}^{1100} \sigma_{T}n_{\rm e}(z)dl
\end{equation}
and
\begin{equation}
{\rm DM}_{\rm IGM}(z)~= \int_{0}^{z}\frac{n_{\rm e}(z^{\prime})}{1+z^{\prime}}dl^{\prime},
\end{equation}
where $\sigma_{T}=6.65\times10^{-25}$cm$^{2}$ is the Thomson scattering cross-section, $dl=cdt=\frac{-cdz}{(1+z)H(z)}$ and $dl^{\prime}=\frac{-cdz^{\prime}}{(1+z^{\prime})H(z^{\prime})}$ are the differentials of the proper distances, and $H(z)$ is the Hubble parameter.
Since the CMB observations measure the integration of $n_{\rm e}$ from the recombination epoch ($z\sim1100$) to $z=0$, solely using $\tau_{\rm CMB}$ is sensitive to only the averaged epoch of reionisation. 
In contrast to $\tau_{\rm CMB}$, FRBs allow us to measure the integration of $n_{\rm e}$ at different redshifts by identifying their host galaxies (see also Section \ref{generate_mock}).
At a fixed redshift, the DM$_{\rm IGM}$ will vary over different lines of sight due to the density fluctuation of intergalactic medium.
For a flat universe, the DM$_{\rm IGM}$ averaged over the different lines of sight is expressed as follows \citep[e.g.,][]{Zhou2014}.
\begin{equation}
\label{eqDMIGM}
\begin{split}
&\langle\mathrm{DM}_{\mathrm{IGM}}\rangle(z)=\Omega_{\mathrm{b}} \frac{3 H_{0} c}{8 \pi G m_{\mathrm{p}}} \times \\
&\int_{0}^{z} \frac{\left(1+z^{\prime}\right) f_{\mathrm{IGM}}\left(z^{\prime}\right)M_{\rm e}\left(z^{\prime} \right)}{\left\{\Omega_{\mathrm{m}}\left(1+z^{\prime}\right)^{3}+\Omega_{\Lambda}\left(1+z^{\prime}\right)^{3\left[1+w\left(z^{\prime}\right)\right]}\right\}^{1 / 2}} d z^{\prime}
\end{split}
\end{equation}
and
\begin{equation}
M_{\rm e}\left(z^{\prime}\right)=Y_{\mathrm{H}} X_{\mathrm{e} \ion{H}{II}}\left(z^{\prime}\right)+ \frac{1}{4}Y_{\rm p}X_{\rm e\ion{He}{II}}\left(z^{\prime}\right) +\frac{1}{2} Y_{\mathrm{p}} X_{\mathrm{e}\ion{He}{III}}\left(z^{\prime}\right),
\end{equation}
where $X_{\rm e\ion{H}{II}}$, $X_{\rm e\ion{He}{II}}$, and $X_{\rm e\ion{He}{III}}$ are the ionisation fractions of intergalactic \ion{H}{II}, \ion{He}{II}, and \ion{He}{III}, respectively.
These are defined as $n_{\ion{H}{II}}/n_{\rm H}$, $n_{\ion{He}{II}}/n_{\rm He}$, and $n_{\ion{He}{III}}/n_{\rm He}$, respectively, where $n_{\rm X}$ represents the number density of element X in the intergalactic medium.
The neutral, singly ionised, and doubly ionised elements are denoted by subscripts I, II, and III, respectively.
$Y_{\mathrm{H}}=\frac{3}{4}$ and $Y_{\mathrm{p}}=\frac{1}{4}$ are the mass fractions of H and He, respectively.
$f_{\mathrm{IGM}}$ is the mass fraction of baryons in the intergalactic medium.
$w(z)$ corresponds to the equation of state of dark energy. 
We assume a constant dark energy, i.e., $w=-1$ \citep{Chevallier2001,Linder2003}.
Following \citet{Zhou2014}, we adopt $f_{\mathrm{IGM}}$ = 0.9 at $z>1.5$ and $f_{\mathrm{IGM}}= 0.053z+0.82$ at $z\leq1.5$ \citep[e.g.,][]{Meiksin2009,Shull2012}.
The parameter $f_{\rm IGM}$ is necessary to take into account the amount of intergalactic ionised material traced by the FRBs.

In this work, we propose to differentiate the  $\langle {\rm DM}_{\rm IGM}\rangle$ against redshift, i.e., $\frac{d\langle{\rm DM_{\rm IGM}}\rangle(z)}{dz}$, to measure $X_{\rm e\ion{H}{II}}$.
At the hydrogen reionisation epoch, $X_{\rm e\ion{He}{II}}$ can be approximated as $X_{\rm e\ion{H}{II}}$ \citep[e.g.,][]{Dai2020} because both neutral hydrogen and neutral helium are singly ionised by ionising photons radiated from star-forming galaxies.
Supposing that the helium reionisation happened at $z \sim 3$-4 \citep[e.g.,][]{Becker2011}, $X_{\rm e\ion{He}{III}} \sim 0$ at $z>4$. 
Therefore, by differentiating Eq. \ref{eqDMIGM} against redshift, $X_{\rm e\ion{H}{II}}$ at the hydrogen reionisation epoch is expressed as
\begin{equation}
\label{eqXe}
\begin{split}
X_{\rm e\ion{H}{II}}(z)=&\frac{d\langle {\rm DM}_{\rm IGM}\rangle (z)}{dz}\frac{3H_{0}c\Omega_{b}}{8\pi G m_{p}}\frac{1}{Y_{\rm H}+\frac{1}{4}Y_{\rm p}} \times\\
&\frac{\left\{\Omega_{\mathrm{m}}\left(1+z\right)^{3}+\Omega_{\Lambda}\left(1+z\right)^{3\left[1+w\left(z\right)\right]}\right\}^{1 / 2}}{(1+z)f_{\rm IGM}(z)}.
\end{split}
\end{equation}
Since $\frac{d\langle {\rm DM}_{\rm IGM}\rangle (z)}{dz}$ will be measured by a slope in the DM$_{\rm IGM}$-$z$ parameter space, Eq. \ref{eqXe} allows us to directly measure $X_{\rm e\ion{H}{II}} (z)$ without any assumption on a functional shape of the reionisation history.
In the following sections, we demonstrate how this method can work in the SKA era using mock FRB data.

\section{Mock FRB data}
\label{mock}
\subsection{Reionisation histories}
To test the method, we consider three different cosmic reionisation histories with the same $\tau_{\rm CMB}$ as prior assumptions.
We describe their parameterisations in the following sections.
The parameters in the three reionisation histories are selected such that $\tau_{\rm CMB}=0.05$, which is calculated using Eq. \ref{eqtauCMB}.
This value is consistent with one of the recent CMB constraints, $\tau_{\rm CMB}=0.054\pm0.007$, within the uncertainty \citep{Planck2020}.
Therefore, solely using $\tau_{\rm CMB}$ is not sensitive to distinguishing between the three reionisation histories presented in this work.
\subsubsection{Tanh reionisation history}
\begin{figure}
    \includegraphics[width=\columnwidth]{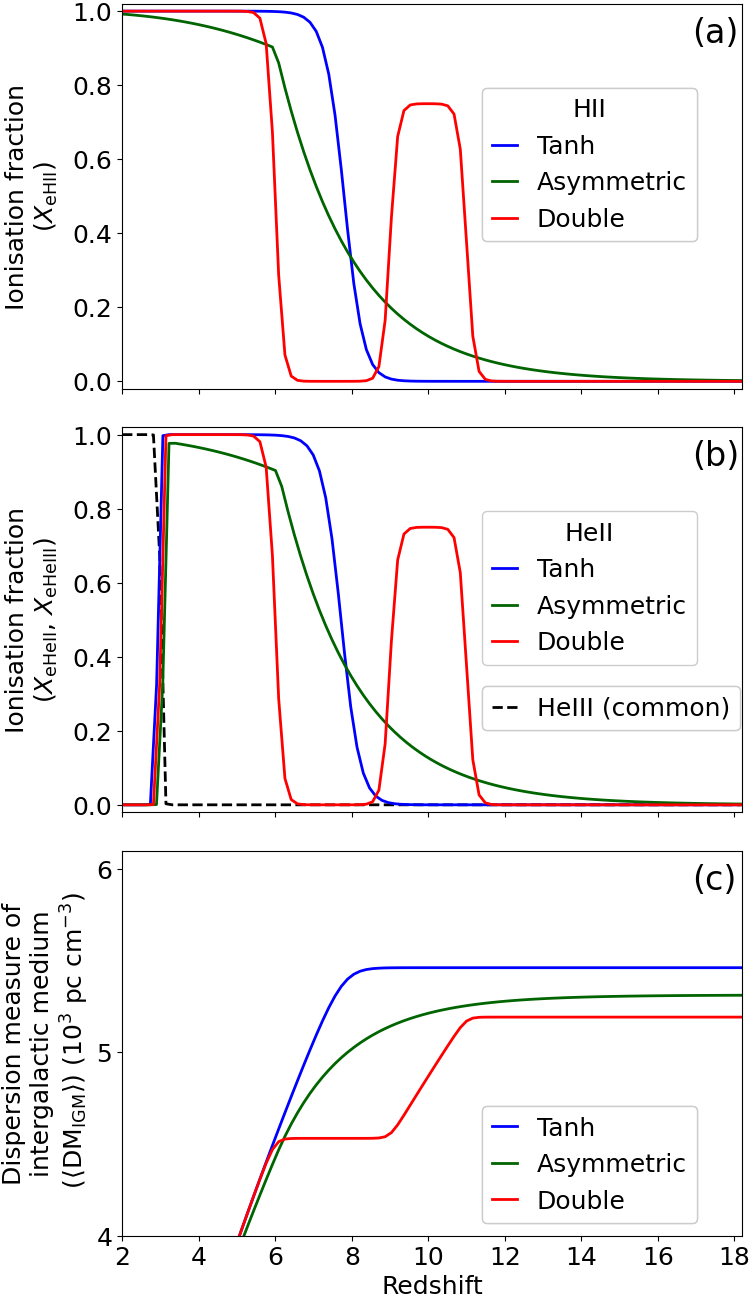}
    \caption{
    (a) Assumed cosmic reionisation histories with the same optical depth of Thomson scattering, $\tau_{\rm CMB}=0.05$ . 
    Their ionisation fractions of hydrogen are shown in different colours.
    (b) Same as top except for ionisation fractions of \ion{He}{II} and \ion{He}{III}.
    The blue and green lines are slightly shifted along the horizontal axis for a visual purpose. 
    The same ionisation fraction of \ion{He}{III} is commonly assumed for the three reionisation histories.
    (c) The dispersion measure of the intergalactic medium as a function of redshift.
    The dispersion measure is averaged over the density fluctuation of the intergalactic medium at each redshift ($\langle {\rm DM}_{\rm IGM}\rangle$).
    The coloured solid lines correspond to the three ionisation histories shown in the panels (a) and (b).
    }
    \label{fig1}
\end{figure}

The first case is a Tanh reionisation. 
This model is conventionally used when the reionisation epoch is constrained by $\tau_{\rm CMB}$ \citep[e.g.,][]{Lewis2008}.
In this model, the ionisation fraction of intergalactic hydrogen, $X_{\rm e,HII}$, is parameterised as follows.
\begin{equation}
\label{Xe_tanh}
X_{\rm e\ion{H}{II}}^{\rm tanh}(z)=1-g(z,z_{\rm re},\Delta z_{\tanh}),
\end{equation}
where
\begin{equation}
\label{eqg}
g(z,z_{\rm re},\Delta z_{\rm tanh})=\frac{1}{2}\left[1+\tanh \left(\frac{(1+z)^{3/2}-(1+z_{\rm re})^{3/2}}{1.5\sqrt{1+z_{\rm re}}\Delta z_{\rm tanh}}\right)\right].
\end{equation}
The parameters, $z_{\rm re}$ and $\Delta z_{\rm tanh}$, represent the redshift at which $X_{\rm e\ion{H}{II}}^{\rm tanh}(z)$ becomes 0.5 (i.e., $X_{\rm e\ion{H}{II}}^{\rm tanh}(z_{\rm re})=0.5$) and the redshift width of the phase transition from neutral to ionised states, respectively.
We adopt $z_{\rm re}=7.8$ and $\Delta z_{\rm tanh}=0.5$.
The model is shown in Fig. \ref{fig1}a in the blue line.

Since both neutral hydrogen and neutral helium are singly ionised by ionising photons from star-forming galaxies, we assume the same functions as the hydrogen reionisation for neutral helium until the \ion{He}{II} is ionised into \ion{He}{III} at the helium reionisation epoch. 
Therefore, the abundance of \ion{He}{II} decreases towards lower redshifts at the helium reionisation epoch, while \ion{He}{III} increases.
The ionisation fractions of \ion{He}{II} are parameterised as follows:
\begin{equation}
\label{Xe_tanh_He}
X_{\rm e\ion{He}{II}}^{\rm tanh}(z)=-g(z,z_{\rm re},\Delta z_{\rm tanh})+g(z,z_{\rm re,He},\Delta z_{\rm He}),
\end{equation}
where $z_{\rm re,He}$ and $\Delta z_{\rm He}$ represent the epoch of helium reionisation and its redshift width, respectively.
We adopt $z_{\rm re, He}=3.0$ and $\Delta z_{\rm He}=0.05$.
The model is shown in Fig. \ref{fig1}b in the blue line.

\subsubsection{Asymmetric reionisation history}
The second case is a redshift-asymmetric model.
Numerically simulated reionisation histories could be parameterised better by the redshift-asymmetric function \citep[][]{Douspis2015}.
This model is parameterised as follows.
\begin{empheq}[left={X_{\rm e\ion{H}{II}}^{\rm asym.}(z)=\empheqlbrace}]{alignat=2}
\label{Xe_asymmetric1}
& 1-\frac{(1-X_{\rm e,6})(1+z)^{3}}{(1+6)^{3}} &(z\leq6)\\
\label{Xe_asymmetric2}
& X_{\rm e, 6} {\rm e}^{\alpha(6-z)} & (z>6),
\end{empheq}
where $X_{\rm e,6}$ is ionisation fraction at $z=6$.
The exponent $\alpha$ is a parameter to describe the duration of transition from neutral to ionised phases.
We adopt $X_{\rm e,H}=0.9$ and $\alpha=0.5$.
The model is shown in Fig. \ref{fig1}a in the green line.

In a similar way to the Tanh reionisation, the ionisation fraction of \ion{He}{II} is approximated as the hydrogen ionisation fraction as follows:
\begin{equation*}
X_{\rm e\ion{He}{II}}^{\rm asym.}(z)=
\end{equation*}
\vspace{-10pt}
\begin{empheq}[left={\empheqlbrace}]{alignat=3}
\label{Xe_asymmetric1_He}
& g(z,z_{\rm re,He},\Delta z_{\rm He}) & (z<3)\\
\label{Xe_asymmetric2_He}
& -\frac{(1-X_{\rm e,6})(1+z)^{3}}{(1+6)^{3}} + g(z,z_{\rm re,He},\Delta z_{\rm He}) &(3\leq z\leq6)\\
\label{Xe_asymmetric3_He}
& X_{\rm e, 6} {\rm e}^{\alpha(6-z)} -\left\{1-g(z,z_{\rm re,He},\Delta z_{\rm He}) \right\}& (z>6).
\end{empheq}
The model is shown in Fig. \ref{fig1}b in the green line.

\subsubsection{Double reionisation history}
The third case is a double reionisation history.
According to reionisation models, the double reionisation processes may have happened at $z\sim10$-15 and $z\sim6$ due to Population III stars and normal star-forming galaxies, respectively \citep[e.g.,][]{Cen2003,Salvador2017}.
The double reionisation at $z>15$ is highly disfavoured by the CMB data \citep[e.g.,][]{Planck2020}.
While the double reionisation is not the main-stream model, it is still useful to demonstrate the feasibility of future FRBs which can reveal detailed substructures of the cosmic reionisation history if such substructures exist.
In this work, we parameterise the double reionisation history using Tanh functions (i.e., Eq \ref{eqg}) as follows:
\begin{equation}
\label{Xe_double}
\begin{split}
X_{\rm e\ion{H}{II}}^{\rm double}(z)=&1+A\left\{-g(z,z_{\rm re1,\Delta z_{\rm double}})+g(z,z_{\rm re2,\Delta z_{\rm double}})\right \}\\
&-g(z,z_{\rm re3,\Delta z_{\rm double}}).
\end{split}
\end{equation}
The redshift parameters, $z_{\rm re1}$, $z_{\rm re2}$, and $z_{\rm re3}$, correspond to the beginning and the end of the first reionisation, and the beginning of the second reionisation, respectively.  
The redshift width of the start and end of reionisation is described by $\Delta z_{\rm double}$.
The maximum ionisation fraction during the first reionisation is parameterised by $A$.
We adopt $z_{\rm re1}=11.0$, $z_{\rm re2}=9.0$, $z_{\rm re3}=6.0$, $\Delta z_{\rm double}=0.2$, and $A=0.75$.
The model is shown in Fig. \ref{fig1}a in the red line.

In a similar way to the Tanh reionisation, the ionisation fraction of \ion{He}{II} is approximated as the hydrogen ionisation fraction as follows:
\begin{equation}
\label{Xe_double_He}
\begin{split}
&X_{\rm e\ion{He}{II}}^{\rm double}(z)= \\
& 1+A\left\{-g(z,z_{\rm re1},\Delta z_{\rm double}) + g(z,z_{\rm re2},\Delta z_{\rm double}) \right\}\\
&-g(z,z_{\rm re3},\Delta z_{\rm double}) + g(z,z_{\rm re,He},\Delta z_{\rm He}).
\end{split}
\end{equation}
The model is shown in Fig. \ref{fig1}b in the red line.

\subsubsection{Helium reionisation}
In this work, a Tanh helium reionisation history is commonly assumed among the three cases of hydrogen reionisation histories.
This is because the helium reionisation has a negligible effect on DM$_{\rm IGM}$ at the hydrogen reionisation epoch which we focus on in this work.
The ionisation fraction of \ion{He}{III} is commonly parameterised as follows:
\begin{equation}
X_{\rm e\ion{He}{III}}^{\rm common}(z)=1-g(z,z_{\rm re,He},\Delta z_{\rm He}).
\end{equation}
This function is shown by the black dashed line in Fig. \ref{fig1}b.

The $\langle {\rm DM}_{\rm IGM} \rangle$ of three reionisation histories are shown in Fig. \ref{fig1}c, which are calculated using Eq. \ref{eqDMIGM}.

\subsection{Number of FRBs to be detected with the SKA}
\begin{figure}
    \includegraphics[width=\columnwidth]{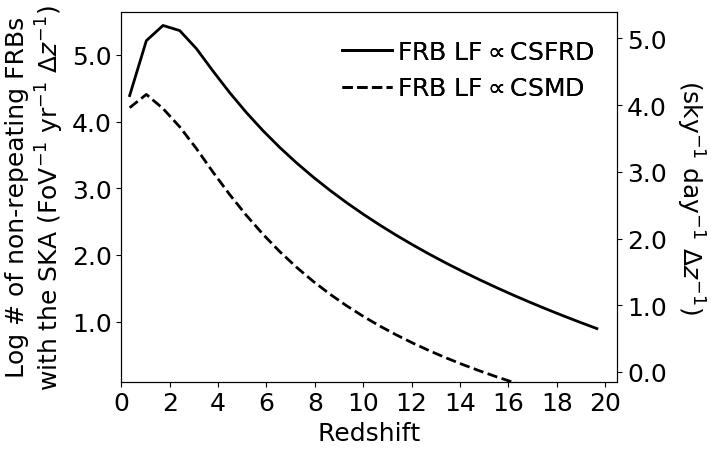}
    \caption{
    Expected number of non-repeating FRBs to be detected with the SKA phase 2 as a function of redshift.
    The number is in units of FoV$^{-1}$ yr$^{-1}$ $\Delta z^{-1}$ (left axis) and sky$^{-1}$ day$^{-1}$ $\Delta z^{-1}$ (right axis), where FoV is the field of view of the SKA Mid frequency aperture array.
    The solid and dashed lines correspond to different assumptions on redshift evolution of FRB luminosity functions (FRB LFs), i.e., the cosmic star formation-rate density (CSFRD) and cosmic stellar-mass density (CSMD), respectively.
    The characteristic spectral index of FRBs is assumed to be $\alpha=-0.3$.
    The observed frequency of SKA is 0.65 GHz.
    The pulse broadening of FRBs by scattering is empirically taken into consideration \citep[see][for details]{Hashimoto2020b}.
    No free-free absorption is assumed in the vicinity of FRB progenitors.
    The adopted Galactic coordinates of FRBs are ($\ell$,$b$)=($45^{\circ}.0$, $-90^{\circ}.0$).
    }
    \label{fig2}
\end{figure}

Recently, the number of FRBs to be detected with the SKA Mid frequency aperture array phase 2 \citep[e.g.,][]{Torchinsky2016} was predicted by \citet{Hashimoto2020b} with a 10 $\sigma$ detection threshold.
While their work was limited up to $z=14$ \citep{Hashimoto2020b}, the first galaxies may appear as early as $z\sim20$ \citep[e.g.,][]{Springel2005,Lacey2011,Bromm2011}.
Therefore, in this work, we extended their calculations towards higher redshifts up to $z=20$ under the same assumptions as follows.
We consider non-repeating FRBs and their luminosity functions which are proportional to either the cosmic stellar-mass density \citep[CSMD;][]{Lopez2018} or cosmic star formation-rate density \citep[CSFRD;][]{Madau2017} \citep[see also][for the redshift evolution of empirically derived FRB luminosity functions]{Hashimoto2020c}.
We use non-repeating FRBs because they are thought to dominate the future FRB sample due to their high number density and brightness \citep{Hashimoto2020b}.
A characteristic spectral index of FRBs is assumed to be $\alpha=-0.3$, which was constrained by broad-band observations between the Green Bank Telescope North Celestial Cap survey at 0.3-0.4 GHz and Parkes surveys at 1.4 GHz \citep{Chawla2017}.
The pulse broadening by scattering is empirically included in the calculations \citep[see][for details]{Hashimoto2020b}.
No free-free absorption is assumed in the vicinity of FRB progenitors in this work.
The adopted Galactic coordinates of FRBs are ($\ell$,$b$)=($45^{\circ}.0$, $-90^{\circ}.0$). 
Note that a small difference in Galactic coordinates negligibly affect the FRB detection \citep{Hashimoto2020b}.
The observed frequency of SKA is centred at 0.65 GHz \citep[e.g.,][]{Torchinsky2016}.
The predicted number of FRBs with the SKA is shown in Fig. \ref{fig2} as a function of redshift.
We note that the 10 sigma detection threshold is utilised for the predictions. 
This threshold is conventionally used for currently detected FRBs. 
Therefore, in this work, we consider high-$z$ FRBs with such high detection significance only.

\subsection{Generating mock data}
\label{generate_mock}
We generate mock data which correspond to non-repeating FRBs to be detected with SKA's one-year observations.
At each redshift bin with $\Delta z=0.67$, mock FRBs are generated such that their number follows Fig. \ref{fig2} for the CSMD and CSFRD cases.
Within each redshift bin, the mock FRBs are distributed uniformly over the redshift.
Their DM$_{\rm IGM}$ are calculated by Eq. \ref{eqDMIGM} assuming the three different cosmic reionisation histories as shown in Fig. \ref{fig1}.
Following the previous works in the literature \citep{Kumar2019b,Linder2020}, the line-of-sight fluctuation of DM$_{\rm IGM}$ ($\sigma_{\rm DM_{\rm IGM}}$) is randomly added to DM$_{\rm IGM}$ of the individual mock FRBs, assuming 
\begin{equation}
\label{DMsigma1}
\sigma_{\rm DM_{\rm IGM}} = 0.2\frac{{\rm DM}_{\rm IGM}}{\sqrt{z}} \hspace{1cm} (z\leq6).
\end{equation}
At $z>6$, $\sigma^{2}_{\rm DM_{\rm IGM}}(z>6)$ is expressed as $\sigma^{2}_{\rm DM_{\rm IGM}}(z=6)+\sigma^{2}_{\rm bubble}(z)$ \citep{Yoshiura2018}, where $\sigma_{\rm bubble}(z)$ is the variance of dispersion measure due to ionised bubbles in the intergalactic medium.
The term $\sigma_{\rm bubble}(z)$ is at most $\sim$ 170 pc cm$^{-3}$ at $z>6$ \citep{Yoshiura2018}.
This value contributes to $\sigma_{\rm DM_{\rm IGM}}(z>6)$ by only $\sim10\%$ under our assumption (Eq. \ref{DMsigma1}). 
Therefore, we approximate $\sigma_{{\rm DM}_{\rm IGM}}(z>6)$ as $\sigma_{{\rm DM}_{\rm IGM}} (z=6)$ in this work.

The randomly added fluctuations of DM$_{\rm IGM}$ follow a Gaussian probability distribution with the standard deviation of $\sigma_{\rm DM_{\rm IGM}}$ at each redshift bin.
We note that DM$_{\rm obs}$ is composed of contributions from the (i) interstellar medium in the Milky Way (DM$_{\rm MW}$), (ii) dark matter halo hosting the Milky Way (DM$_{\rm halo}$), (iii) intervening galaxies (DM$_{\rm inter,gal}$), (iv) intergalactic medium (DM$_{\rm IGM}$), and (v) galaxy hosting the FRB (DM$_{\rm host}$).
The DM$_{\rm inter,gal}$ has only a minor effect on DM$_{\rm obs}$ \citep[DM$_{\rm inter,gal}\lesssim0.1$ pc cm$^{-3}$ on average at all $z<7$;][]{Prochaska2018}.
Therefore, the DM$_{\rm IGM}$ is measured by subtracting DM$_{\rm MW}$, DM$_{\rm halo}$, and DM$_{\rm host}$ from DM$_{\rm obs}$ \citep[e.g.,][]{Hashimoto2020c}.
While the DM$_{\rm halo}$ and DM$_{\rm host}$ are poorly understood, their uncertainties (order of 10 pc cm$^{-3}$) are much smaller than the DM$_{\rm IGM}$ component (order of 10$^{3}$ pc cm$^{-3}$) and its line-of-sight fluctuation (order of $10^{2}$ pc cm$^{-3}$) at the reionisation epoch \citep[e.g.,][]{Prochaska2019DMhalo}.
Therefore, we neglect the uncertainties of DM$_{\rm halo}$ and DM$_{\rm host}$ subtractions from the DM$_{\rm obs}$ to derive the DM$_{\rm IGM}$ observationally.

In terms of the DM$_{\rm obs}$ uncertainty, the difference between currently detected low-$z$ FRBs ($z\lesssim2$) and future high-$z$ ones would be their slopes in the frequency-time space. 
The uncertainty of the slope corresponds to the DM$_{\rm obs}$ uncertainty.
For instance, Parkes detected FRBs up to $z\sim2$ \citep[e.g.,][]{Hashimoto2020c} with a spectral resolution of $\sim$400 kHz \citep[e.g.,][]{Keane2015}. 
The slope becomes $\sim$3 times flatter from $z\sim2$ to $z\sim20$. 
However, the SKA’s spectral resolution is $\sim$10kHz \citep{Torchinsky2016} which is $\gtrsim$10 times better than the 400 kHz spectral resolution of Parkes.
Therefore, the SKA is expected to sample the slopes of high-$z$ FRBs as accurately as the current FRB observations or even better.
A typical observational uncertainty of DM$_{\rm obs}$ is $\lesssim 1$\% using current radio telescopes \citep[e.g.,][]{Petroff2016}.
Therefore, we can reasonably ignore the DM$_{\rm obs}$ uncertainty of mock FRBs in the SKA era.

We consider three cases of redshift measurements of FRBs: (A) spectroscopic redshift available for 100\% of the FRB sample, (B) spectroscopic redshift available for 10\% of the FRB sample, and (C) redshift estimated from an empirical relation between the time-integrated luminosity and duration of non-repeating FRBs.
Each case assumes two different redshift evolutions of FRB luminosity functions (CSMD and CSFRD) and three different reionisation histories as mentioned above.

\subsubsection{Mock data A: spectroscopic redshift for 100\% sample}
\label{mocka}
\begin{figure}
    \includegraphics[width=\columnwidth]{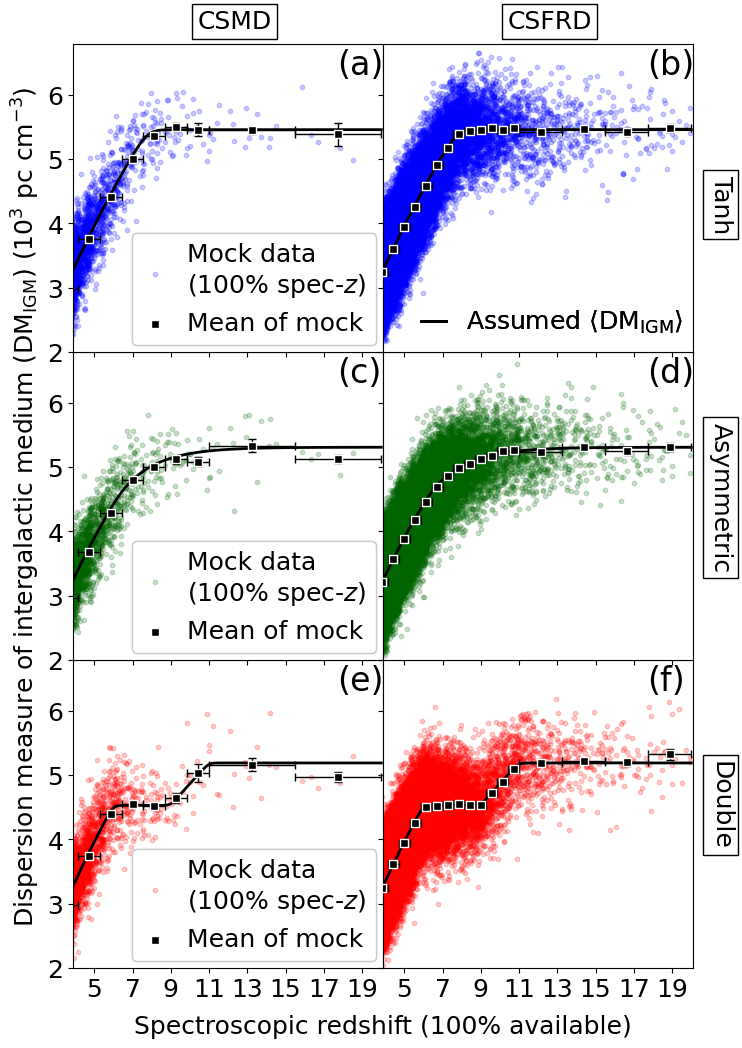}
    \caption{
    Simulated mock FRB data to be detected with the Square Kilometre Array (Mock data A).
    We assumed that spectroscopic redshifts are available for all of the mock data.
    The redshift uncertainty is ignored.
    From left to right panels, redshift evolutions of FRB luminosity functions are assumed to be proportional to the cosmic stellar-mass density (CSMD) and star formation-rate density (CSFRD), respectively.
    From top to bottom panels, Tanh, asymmetric, and double reionisation histories are assumed, respectively.
    The black squares indicate averaged dispersion measures of mock FRB data at redshift bins, $\langle {\rm DM}_{\rm IGM, mock}\rangle(z)$.
    The redshift bins are shown by horizontal bars.
    The vertical error bars indicate uncertainties of $\langle{\rm DM}_{\rm IGM, mock}\rangle(z)$.
    These uncertainties are calculated by standard errors of mock data in the redshift bins, i.e., $\sigma_{\rm DM_{\rm IGM, mock}}/\sqrt{N_{\rm mock}}$, where $\sigma_{\rm DM_{\rm IGM, mock}}$ and $\sqrt{N_{\rm mock}}$ are the standard deviation of DM$_{\rm IGM,mock}$ and number of mock FRB data in each redshift bin.
    }
    \label{fig3}
\end{figure}

The spectroscopic redshifts of FRBs can be measured by identifying their host galaxies \citep[e.g.,][]{Macquart2020} at FRBs' positions localised by the SKA \citep{Torchinsky2016}.
We assume one-year FRB detection with the SKA in this work. 
In contrast, the spectroscopic follow-up observations of FRB host galaxies can be performed over decades using future telescopes such as the European Extremely Large Telescope \citep{Gilmozzi2007} and the James Webb Space Telescope \citep{Gardner2006}. 
The SKA may also be able to measure spectroscopic redshifts of host galaxies at $z\sim10$ by detecting the 21-cm emission line \citep[e.g.,][]{Ghara2016}.
Therefore, future extensive efforts to measure spectroscopic redshifts of the host galaxies may significantly increase the number of FRBs with spectroscopic redshift. 
The number of such FRBs may approach the number of one-year FRB detection with the SKA.
This case corresponds to the 100\% availability of the spectroscopic redshifts for one-year of FRB detection with the SKA.
This is an optimistic assumption.
However, it is useful for demonstrating the potential of FRBs in investigating the cosmic reionisation history.
Therefore, we show this case as the most ideal case.
The redshift uncertainties of mock FRBs are approximated as $\sigma_{z}=0$ in this case.
The generated mock FRB data (hereafter \lq Mock data A\rq) are shown in Fig. \ref{fig3}.

\subsubsection{Mock data B: spectroscopic redshift for 10\% sample}
\label{mockb}
\begin{figure}
    \includegraphics[width=\columnwidth]{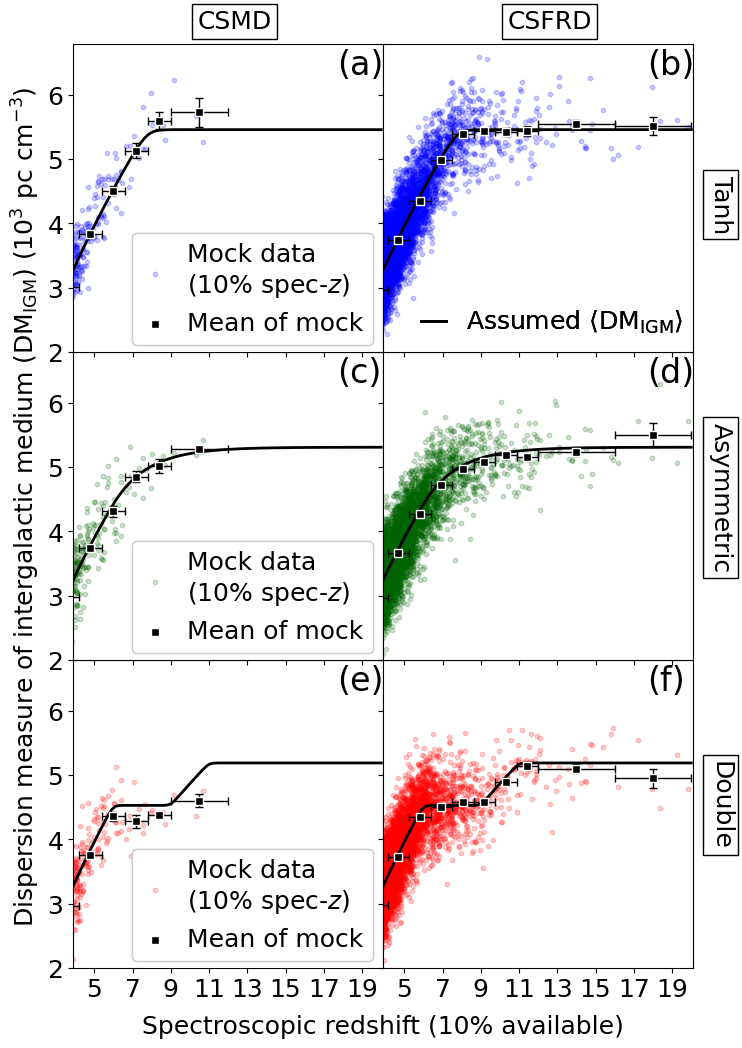}
    \caption{
    Same as Fig. \ref{fig3} except that the spectroscopic redshift is available for 10\% of the mock FRB data (Mock data B).
    }
    \label{fig4}
\end{figure}

FRB host identifications and their spectroscopic follow-up observations are expensive and time-consuming in general.
Therefore, the spectroscopic redshifts will likely be available only for a part of the FRB sample.
Currently, $\sim$10\% of FRBs have their host identifications and spectroscopic redshifts\footnote[1]{\url{http://frbhosts.org/}}\citep{Heintz2020}.
For this mock data, we assume that the spectroscopic redshifts are available for 10\% of the FRBs detected with the SKA in one year.
An analogy is the host identification for LGRBs.
So far $\sim$ 2,000 LGRBs have been localised down to arcmin level \footnote[2]{\url{https://www.mpe.mpg.de/~jcg/grbgen.html}}.
Among them, $\sim$10\% have their host identifications and spectroscopic redshifts\footnote[3]{\url{http://www.grbhosts.org/}}.
The SKA will achieve better localisations for FRBs down to arcsec level \citep{Torchinsky2016}. 
It is conservative to assume spectroscopic redshifts for 10\% of the future FRBs.
In this case, mock data include 10\% of the originally generated FRBs with accurate redshift measurements, i.e., $\sigma_{z}=0$.
The generated mock FRB data (hereafter \lq Mock data B\rq) are shown in Fig. \ref{fig4}.

\subsubsection{Mock data C: Redshift estimated from an empirical relation}
\label{mockc}
\citet{Hashimoto2019} reported an empirical relationship between the time-integrated luminosity ($L_{\nu}$) and rest-frame intrinsic duration ($w_{\rm int, rest}$) of non-repeating FRBs (hereafter, $L_{\nu}$-$w_{\rm int, rest}$ relation).
The $L_{\nu}$-$w_{\rm int, rest}$ relation can be used to constrain redshifts of individual FRBs without the host identification.
The observed dispersion around the relation is 0.28 dex in $\log(w_{\rm int, rest})$ axis.
This dispersion is dominated by the observational uncertainties of the currently detected FRBs \citep{Hashimoto2019}.
Since the observational uncertainties of SKA will be $\sim$ two orders of magnitude smaller than that of currently working radio telescopes such as Parkes, the dispersion around the $L_{\nu}$-$w_{\rm int, rest}$ relation would be significantly reduced from the current measurement of 0.28 dex.
However, the dispersion may still be dominated by the intrinsic variation of FRBs rather than the observational uncertainty.
In this work, we assume 10\% of the dispersion estimated for the current FRB sample.
The 10\% dispersion of 0.28 dex in $\log(w_{\rm int, rest})$ corresponds to a 0.13 dex dispersion in the log time-integrated luminosity \citep{Hashimoto2019}.
This 0.13 dex dispersion is twice as large as the luminosity dispersion of type Ia supernovae since the dispersion of intrinsic luminosity of type Ia supernovae is $\sim$ 0.15 mag, i.e., 0.15/2.5=0.06 dex in log luminosity scale \citep[e.g.,][]{Mohlabeng2014,Pan2014}.

\begin{figure}
    \includegraphics[width=\columnwidth]{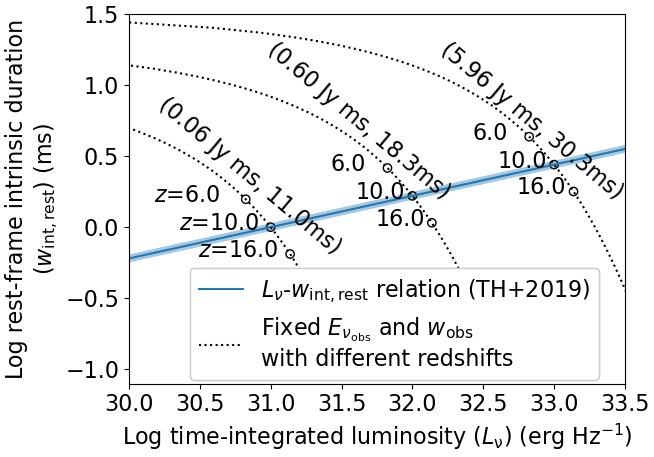}
    \caption{
    An empirical relation between the time-integrated luminosity and rest-frame intrinsic duration of FRBs ($L_{\nu}$-$w_{\rm int, rest}$ relation).
    The assumed dispersion of $L_{\nu}$-$w_{\rm int, rest}$ relation in the SKA era is indicated by a shaded blue region, which is 10\% of the observed dispersion estimated from currently detected FRBs \citep{Hashimoto2019}.
    The dotted lines indicate redshift tracks of FRBs with fixed observed fluences ($E_{\nu_{\rm obs}}$) and durations ($w_{\rm obs}$) calculated from Eq. 6 in \citet{Hashimoto2019} and pulse broadening effect by redshift.
    The open circles labelled with redshift values correspond to $z=6.0, 10.0$ and $16.0$ on the tracks.
    The intersection between the $L_{\nu}$-$w_{\rm int, rest}$ relation and dotted track provide each FRB with a redshift estimate.
    The presented $E_{\nu_{\rm obs}}$ and $w_{\rm obs}$ are examples of FRBs at $z=10$.
    The instrumental pulse broadening by the SKA is ignored since it is much smaller than the redshift broadening effect at the hydrogen reionisation epoch.
    }
    \label{fig5}
\end{figure}

\begin{figure}
    \includegraphics[width=\columnwidth]{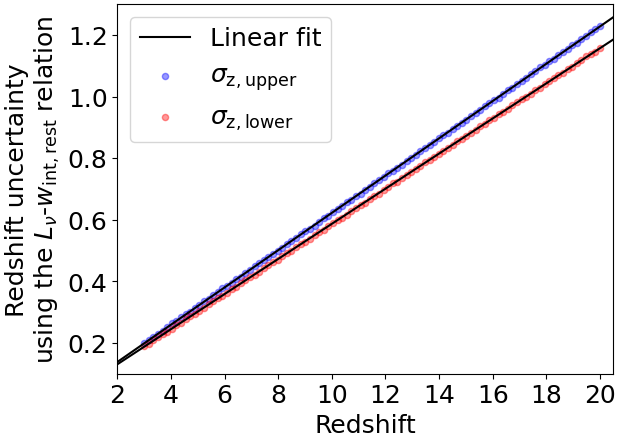}
    \caption{
    The redshift uncertainty of FRBs as a function of redshift using the empirical relation between the time-integrated luminosity and rest-frame intrinsic duration ($L_{\nu}$-$w_{\rm int, rest}$ relation).
    The blue and red dots represent upper and lower bounds of redshift uncertainties, respectively. 
    These data points are calculated from overlapping regions between the dispersion of $L_{\nu}$-$w_{\rm int, rest}$ relation and redshift tracks in Fig. \ref{fig5}.
    The solid lines indicate linear fits to the upper and lower bounds as a function of redshift.
    }
    \label{fig6}
\end{figure}

\begin{figure}
    \includegraphics[width=\columnwidth]{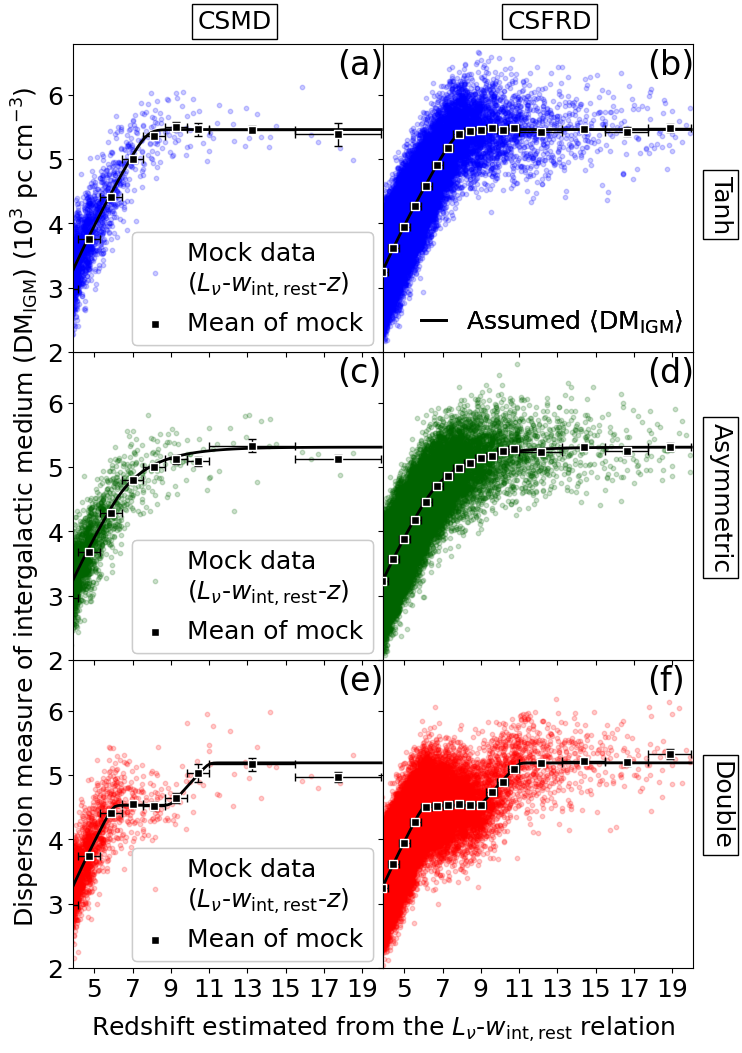}
    \caption{
    Same as Fig. \ref{fig3} except that redshift uncertainties are randomly added to individual data using Eqs. \ref{eqsigmau} and \ref{eqsigmal} (Mock data C).
    }
    \label{fig7}
\end{figure}

The assumed $L_{\nu}$-$w_{\rm int, rest}$ relation and its dispersion are shown in Fig. \ref{fig5}.
The dotted lines in Fig. \ref{fig5} indicate redshift tracks of FRBs with fixed observed fluences and durations.
The redshift tracks are calculated from Eq. 6 in \citet{Hashimoto2019} and pulse broadening effect by redshift.
Therefore, the intersection between the $L_{\nu}$-$w_{\rm int, rest}$ relation and redshift track provides each FRB with a redshift estimate.
The instrumental pulse broadening by the SKA is $\sim$1 ms for an FRB with DM$_{\rm obs}=5\times10^{3}$ pc cm$^{-1}$ \citep{Hashimoto2020b}.
This value is negligible compared to the redshift broadening effect at the hydrogen reionisation epoch (e.g., 11 ms for an FRB with $w_{\rm int,rest}=1.0$ ms at $z=10$).
The redshift uncertainty is calculated as a function of redshift using the overlapping region between the dispersion of $L_{\nu}$-$w_{\rm int, rest}$ relation (blue shaded region) and redshift tracks (dotted lines).
The upper and lower bounds of redshift uncertainties ($\sigma_{\rm z,upper}$ and $\sigma_{\rm z,lower}$, respectively) are shown in Fig. \ref{fig6}. 
The best-fit linear functions to the uncertainties are
\begin{equation}
\label{eqsigmau}
\sigma_{\rm z,upper}=0.061z+0.017
\end{equation}
and
\begin{equation}
\label{eqsigmal}
\sigma_{\rm z,lower}=0.057z+0.017.
\end{equation}
Using Eqs. \ref{eqsigmau} and \ref{eqsigmal}, we randomly added redshift uncertainties to the mock FRB data.
At each redshift, the randomised redshift errors follow an asymmetric Gaussian probability distribution which has the standard deviation of $\sigma_{\rm z,upper}$ for one side and $\sigma_{\rm z,lower}$ for the other side.
The generated mock FRB data (hereafter \lq Mock data C\rq) are shown in Fig. \ref{fig7}.

\section{Results}
\label{results}
\begin{figure*}
    \includegraphics[width=2.0\columnwidth]{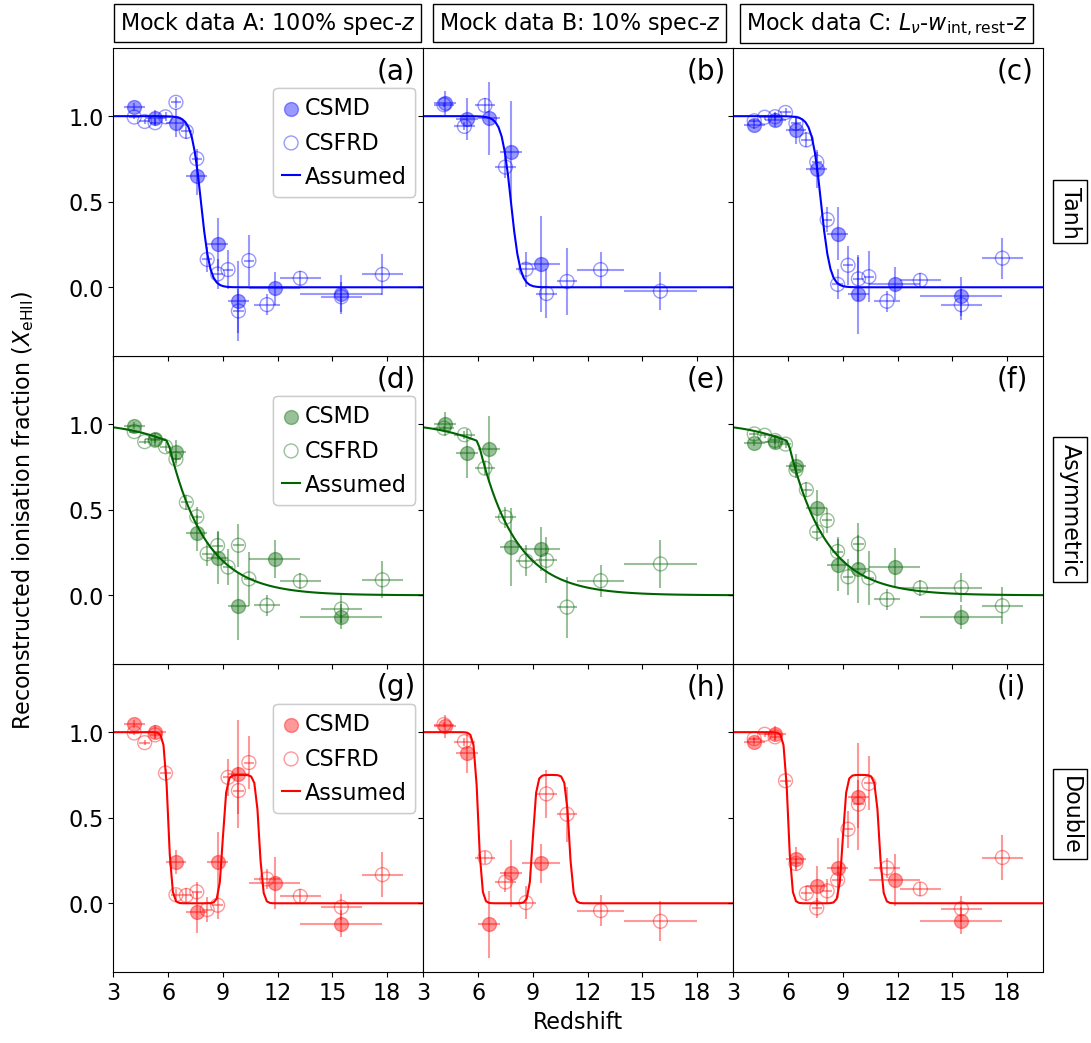}
    \caption{
    Ionisation fractions as a function of redshift, $X_{\rm e\ion{H}{II}}(z)$, reconstructed from mock FRB data.
    The reconstructed data are shown by filled and open circles. 
    The filled and open circles correspond to assumptions on redshift evolutions of FRB luminosity functions which are proportional to the cosmic stellar-mass density (CSMD) and cosmic star formation-rate density (CSFRD), respectively.
    The horizontal bars indicate redshift bins.
    The vertical bars indicate uncertainties of $X_{\rm e\ion{H}{II}}$ which include uncertainties of $\langle {\rm DM}_{\rm IGM, mock}\rangle(z)$ in Figs. \ref{fig3}, \ref{fig4}, and \ref{fig7}.
    The prior assumptions on reionisation histories are shown by solid lines.
    From left to right panels, Mock data A (spectroscopic redshift available for 100\% of FRB sample), B (spectroscopic redshift available for 10\% of FRB sample), and C (redshift estimated from an empirical relation between the time-integrated luminosity and rest-frame intrinsic duration of non-repeating FRBs: $L_{\nu}$-$w_{\rm int, rest}$ relation) are shown, respectively.
    From top to bottom panels, Tanh, asymmetric, and double reionisation histories are used as prior assumptions, respectively.
    }
    \label{fig8}
\end{figure*}

Here we reconstruct $X_{\rm e\ion{H}{II}}(z)$ from the mock FRB data described in Section \ref{mock}.
For this purpose, we calculated averaged DM$_{\rm IGM}$ of mock FRB data, $\langle{\rm DM}_{\rm IGM,mock}\rangle(z)$, at every redshift bin.
The redshift bins are flexibly determined depending on the number of mock data.
Mock data A has 7 (14) and 2 (4) bins at $3<z\leq11$ and at $11<z\leq20$, respectively, for the CSMD (CSFRD) case.
Mock data B has 5 (8) and 1 (2) bins at $3<z\leq9$ ($3<z\leq12$) and at $9<z\leq12$ ($12<z\leq20$), respectively, for the CSMD (CSFRD) case.
The redshift bins of Mock data C are the same as that of Mock data A.
$\langle{\rm DM}_{\rm IGM,mock}\rangle(z)$ are shown by black squares in Figs. \ref{fig3}, \ref{fig4}, and \ref{fig7}.
Their uncertainties are calculated by standard errors of mock FRB data in the redshift bins, i.e., $\sigma_{\rm DM_{\rm IGM, mock}}/\sqrt{N_{\rm mock}}$, where $\sigma_{\rm DM_{\rm IGM, mock}}$ and $\sqrt{N_{\rm mock}}$ are the standard deviation of DM$_{\rm IGM,mock}$ and the number of mock FRB data in each redshift bin.
The measured $\langle{\rm DM}_{\rm IGM,mock}\rangle(z)$ was used to estimate the slopes of the mock FRB data, $\frac{d\langle{\rm DM}_{\rm IGM,mock}\rangle(z)}{dz}$, in Figs. \ref{fig3}, \ref{fig4}, and \ref{fig7}.
This slope measurement allows us to reconstruct $X_{\rm e\ion{H}{II}}(z)$ using Eq. \ref{eqXe}.
The reconstructed $X_{\rm e\ion{H}{II}}(z)$ are shown in Fig. \ref{fig8}.
The three reionisation histories and three cases of redshift measurements (Mock data A, B, and C) correspond to nine panels in Fig. \ref{fig8}.
The dominant source of the errors in Fig. \ref{fig8} is the uncertainty of $\langle {\rm DM}_{\rm IGM}\rangle$ in each redshift bin.
This uncertainty is calculated by a standard error in each redshift bin in Figs. \ref{fig3}, \ref{fig4}, and \ref{fig7} as mentioned above.
In all of the cases, the assumed reionisation histories are reasonably reproduced.
We did not assume any functional shape of the reionisation history when $X_{\rm e\ion{H}{II}}$ was reconstructed from the mock FRB data.
Even the shape of double reionisation can be reconstructed (bottom panels in Fig. \ref{fig8}), which indicates the capability of future FRBs and the feasibility of our method.

\section{Discussion}
\label{discussions}
\subsection{Reionisation history reconstructed from mock FRB data}
The CMB observations constrain $\tau_{\rm CMB}$ \citep[e.g.,][]{Pagano2020}.
However, the ionisation fraction as a function of redshift is difficult to be measured solely by using $\tau_{\rm CMB}$ since $\tau_{\rm CMB}$ is an integrated value over redshift $z=0$ to $\sim$1100.
In this sense, solely using $\tau_{\rm CMB}$ is not sensitive to distinguishing between the three reionisation histories with the same $\tau_{\rm CMB}$ value presented in this work.
We note that the detailed redshift evolution of the ionisation fraction is imprinted in the shape of the power spectrum of CMB polarisation \citep[][]{Zaldarriaga1997,Kaplinghat2003,Planck2020}.
Therefore, the CMB polarisation can constrain the high-$z$ end of reionisation \citep[e.g.,][]{Planck2020}.

The DM$_{\rm IGM}$ of FRBs is a similar parameter to $\tau_{\rm CMB}$.
In previous studies \citep[e.g.,][]{Ioka2003,Inoue2004,Fialkov2016}, the DM$_{\rm IGM}$ has been proposed as a tool to constrain the reionisation history independently from the $\tau_{\rm CMB}$ measurement.
\citet{Fialkov2016} showed that different reionisation histories give rise to different shapes of DM$_{\rm IGM}$ as a function of redshift.
Therefore, having a large sample of FRBs would improve the constraint on the redshift evolution of the cosmic ionised fraction.
However, $d\langle{\rm DM_{\rm IGM}}\rangle/dz$ has been rarely discussed in spite of its better sensitivity to the reionisation history than DM$_{\rm IGM}(z)$ itself \citep{Zheng2014}.
In this work, we emphasise that the advantage of FRBs over CMB is the availability of DM$_{\rm IGM}$ as a function of redshift, i.e., availability of $d\langle{\rm DM_{\rm IGM}}\rangle/dz$.
By taking advantage of FRBs, we will be able to directly measure the ionisation history as a function of redshift.
Fig. \ref{fig8} indicates that future FRBs are able to distinguish between these histories using our method.
Having only 10\% of the spectroscopic redshift measurements of the SKA FRBs might be enough to reveal the cosmic reionisation history.

Recently, \citet[][]{Dai2020} proposed to use an auto-correlation power spectrum of DM$_{\rm IGM}$ measurements to constrain the reionisation epoch.
Based on their mock FRB data, they predicted that the duration of the transition phase of reionisation will be significantly constrained compared to using the CMB data only.
The advantage of this method is that accurate redshift measurements are not necessary but a rough redshift distribution is enough for the analysis.
However, their method has to assume a functional shape of the reionisation history which is poorly known.
On the other hand, our method requires the redshift measurements for individual FRBs, but it is able to measure the cosmic reionisation history as it is without the shape assumption.
Motivated by \citet{Dai2020}, redshift information of FRBs may be obtained by a spatial cross-correlation between FRBs and high-$z$ galaxies whose redshifts are known from galaxy surveys (e.g., Hsiao et al. 2021 in prep.).
This may provide FRBs at each DM$_{\rm IGM}$ bin with a representative redshift, which will work in a similar way to our method.

\subsection{Assumptions in the luminosity-duration relation}
The empirically derived $L_{\nu}$-$w_{\rm int, rest}$ relation of non-repeating FRBs \citep{Hashimoto2019} is used to estimate the redshifts of Mock data C. 
We caution that there are two assumptions behind Mock data C.

One is the reduction of dispersion around the relation in the SKA era.
As we mentioned in Section \ref{mockc}, the dispersion of the relation is dominated by the observational uncertainties for the current FRB sample \citep{Hashimoto2019}.
The observational uncertainties will decrease significantly for the FRBs detected with the SKA phase 2 \citep[e.g.,][]{Torchinsky2016}.
The dispersion around the relation would be dominated by the intrinsic variation of FRBs in the SKA era.
Indeed some FRB models predict positive correlations between the time-integrated luminosity and duration \citep[e.g.,][]{Lyubarsky2014,Geng2015,Romero2016}.
In these models, the time-integrated luminosity depends not only on the duration but also on other physical parameters.
Therefore, one possible way to reduce the intrinsic dispersion is by introducing (an) additional factor(s) to the relation. 
In the past, the peak luminosity of type Ia supernovae has been reported with a dispersion of 0.8 mag in the $B$ band \citep{Phillips1993}.
So far, extensive efforts were made to reduce the dispersion by taking into account new factors such as light-curve shape, colour, and host properties of type Ia supernovae \citep[e.g.,][]{Pan2014}. 
Such corrections for the relation allowed us to utilise type Ia supernovae as a standard candle eventually.
Similarly, substantial efforts will be necessary for the redshift estimate using the $L_{\nu}$-$w_{\rm int, rest}$ relation.

Another assumption is the \lq no redshift evolution\rq\ of the $L_{\nu}$-$w_{\rm int, rest}$ relation.
If the relation depends on redshift, the redshift measured from the $L_{\nu}$-$w_{\rm int, rest}$ relation contains a systematic uncertainty.
Unless the relation is corrected for the evolution, the derived reionisation history will systematically shift towards lower/higher redshifts while the shape of the ionisation history may be less affected.
For the redshift-measurement purpose, the $L_{\nu}$-$w_{\rm int, rest}$ relation also needs to be well established as a function of redshift.

\section{Conclusion}
\label{conclusions}
We propose a method to directly measure the ionisation fraction of the intergalactic medium as a function of redshift using fast radio bursts (FRBs).
The dispersion measure in the intergalactic medium (DM$_{\rm IGM}$) derived from FRBs is unique because it enables us to measure the integrated electron densities in the intergalactic medium at different redshifts.
The differential of averaged DM$_{\rm IGM}$ against redshift, $d\langle{\rm DM_{\rm IGM}}\rangle/dz$, is proportional to the ionisation fraction. 
Therefore, $d\langle{\rm DM_{\rm IGM}}\rangle/dz$ allows us to directly measure the ionisation fraction of the intergalactic medium as a function of redshift.
We consider future non-repeating FRB sources to be detected with the Square Kilometre Array (SKA) at the cosmic reionisation epoch.
For a demonstration purpose, three cosmic ionisation histories including Tanh, asymmetric, and double reionisation are assumed as priors.
Since these histories are parameterised with the same Thomson-scattering optical depth ($\tau_{\rm CMB}$), solely using $\tau_{\rm CMB}$ constrained from the cosmic microwave background data is not sensitive to distinguishing between these histories.
Based on mock FRB data in the SKA era, we found that the cosmic reionisation history can be reasonably reconstructed by using $d\langle{\rm DM_{\rm IGM}}\rangle/dz$ for each reionisation case.
In the reconstruction process, no assumption is made on a functional shape of the cosmic reionisation history.
The advantage of our method is that it allows us to measure the reionisation history as it is, in contrast to previously proposed methods solely using $\tau_{\rm CMB}$ or DM$_{\rm IGM}$.
Our results indicate the capability of future FRBs and the feasibility of our method in revealing the cosmic reionisation history.

\section*{Acknowledgements}
We are very grateful to the anonymous referee for many insightful comments.
We thank Dr. Susumu Inoue and Dr. Shintaro Yoshiura for useful discussions.
TH and AYLO are supported by the Centre for Informatics and Computation in Astronomy (CICA) at National Tsing Hua University (NTHU) through a grant from the Ministry of Education of Taiwan.
TG acknowledges the support by the Ministry of Science and Technology of Taiwan through grant 108-2628-M-007-004-MY3.
AYLO's visit to NTHU was supported by the Ministry of Science and Technology of the ROC (Taiwan) grant 105-2119-M-007-028-MY3, hosted by Prof. Albert Kong.
This work used high-performance computing facilities operated by the CICA at NTHU. 
This equipment was funded by the Ministry of Education of Taiwan, the Ministry of Science and Technology of Taiwan, and NTHU.
This research has made use of NASA's Astrophysics Data System.

\section*{Data availability}
The data underlying this article will be shared on reasonable request to the corresponding author.


\bibliographystyle{mnras}
\bibliography{Reionisation_mnras} 




\bsp	
\label{lastpage}
\end{document}